\documentclass[printer]{aa}
\usepackage[utf8]{inputenc}
\usepackage{graphicx}
\usepackage{txfonts}
\usepackage{tabularx}
\usepackage{adjustbox}
\usepackage[normalem]{ulem}
\usepackage[flushleft]{threeparttable}
\usepackage[normalem]{ulem}
\def\gaia{\textit{Gaia}\xspace}
\def\gmag{$G$\xspace}
\def\gbp{$G_{\rm BP}$\xspace}
\def\grp{$G_{\rm RP}$\xspace}

\newcommand{\rrl}{RR~Lyrae\xspace}
\def\parallax{$\varpi$\xspace}
\providecommand{\sigparallax}{\ensuremath{\sigma_{\varpi}}\xspace}

\begin{document}

\title{\gaia Data Release 3:}
\subtitle{Ellipsoidal Variables with Possible Black-Hole or Neutron Star secondaries}
\titlerunning{Ellipsoidal Variables with Possible Compact Companions}
\author{R. Gomel\inst{1} \and T. Mazeh\inst{1} \and S. Faigler\inst{1} 
\and D. Bashi\inst{1}
\and L. Eyer\inst{2} \and L. Rimoldini\inst{3} \and M. Audard \inst{2}
\and N. Mowlavi \inst{2,3}
\and B. Holl \inst{3} \and G. Jevardat \inst{3} \and K. Nienartowicz  \inst{3}
\and I. Lecoeur \inst{3}
\and L. Wyrzykowski\inst{4}
}
\institute{School of Physics and Astronomy, Tel Aviv University, Tel Aviv, 6997801, Israel\\ 
\email{roygomel@tauex.tau.ac.il}
\and 
Department of Astronomy, University of Geneva, Chemin Pegasi 51, CH-1290 Versoix, Switzerland
\and
Department of Astronomy, University of Geneva, Chemin d’Ecogia 16, CH-1290 Versoix, Switzerland
\and
Warsaw University Astronomical Observatory
Department of Physics
Al. Ujazdowskie 4
00-478 Warszawa
Poland
}
\date{June 2022}
\abstract{As part of \gaia Data Release 3, supervised classification identified a large number of ellipsoidal variables, for which the periodic variability is presumably induced by tidal interaction with a companion in a close binary system. In this paper, we present $6306$ short-period probable ellipsoidal variables with relatively large-amplitude \gaia \gmag-band photometric modulations, indicating a possible massive, unseen secondary. In case of a main-sequence primary, the more massive secondary is probably a compact object --- either a black hole or a neutron star, and sometimes a white dwarf. The identification is based on a robust {\it modified} minimum mass ratio (mMMR) suggested recently by Gomel, Faigler and Mazeh (2021), derived from the observed ellipsoidal amplitude only, without the use of the primary mass or radius. We also list a subset of $262$ systems with mMMR larger than unity, for which the compact-secondary probability is higher. Follow-up observations are needed to verify the true nature of these variables.}
\maketitle
%
\section{Introduction}

About 20 dynamically confirmed Galactic stellar black holes (BH) 
are known to reside in close binary systems. 
All known stellar-mass BHs have been discovered by their X-ray emission, fueled by mass transfer from their non-compact stellar companions
\citep[e.g.,][]{fabian89, remillard06, orosz07, ziolkowski14}.
However, a large fraction of BHs in binaries were probably not detected yet
because their optical counterparts are well within their Roche lobes, so mass is not transferred and X-rays are not generated, making these systems dormant BHs 
\citep[see discussion on the frequency of such systems by][]{breivik17, mashian17, yamaguchi18, shao19, yi19, wiktorowicz19, shikauchi20}.
Similar arguments apply to dormant neutron stars (NS). Only very few known X-ray binaries harbor a NS companion, while most of similar binaries are probably dormant.

Such a dormant binary can be identified 
either by a large radial-velocity modulation \citep[e.g.,][]{thompson19,Zheng19,clavel21,jayasinghe22}, or
by the stellar large ellipsoidal modulation, induced by tidal interaction with its compact companion \citep[e.g.,][]{rowan21}.
We are aiming to identify some of the short-period dormant compact systems by the ellipsoidal effect, as revealed by the \gaia photometry. A similar study was performed by
\citet{gomel21c} 
for the ellipsoidals identified by the OGLE team toward the Galactic bulge \citep{soszy16}.

Based on the observed ellipsoidal amplitude and the estimated mass and radius of the primary star, one can derive a minimum secondary-to-primary mass ratio, 
defined as the mass ratio obtained for an inclination of $90^{\circ}$,
 provided most of the light is coming from the primary star \citep[e.g.,][]{faigler11,faigler15}.
A binary with a minimum mass ratio significantly larger than unity might be a candidate for having a dormant compact-object companion --- a BH, a NS, or even a white dwarf, depending on the estimated minimum mass of the secondary. 

Unfortunately, in many cases the primary mass and radius are not well known.
Therefore, \citet*{gomel21b} 
presented a simple approach that circumvents this problem by suggesting a robust {\it modified} minimum mass ratio (mMMR), assuming the primary fills its Roche lobe. 
The newly defined mMMR depends on the ellipsoidal amplitude, and to some extent (on the order of a few percent) on the primary effective temperature, but does not depend on the primary-star mass or radius. 

The mMMR is always {\it smaller} than the minimum mass ratio, which is, in its turn, smaller than the actual mass ratio. Therefore, binaries with a large mMMR are good candidates for having a compact-object secondary,
even if we cannot reliably constrain their primary mass and radius. This is specifically true for binaries with mMMR larger than unity, provided the periodic modulation is due to the ellipsoidal effect.

In this paper, we analysed tens of millions of systems that were classified as possible ellipsoidals, with varying degrees of
certainty,
by \gaia DR3 photometric pipeline. We apply our compact-companion software to analyze the \gaia light curves of these stars, searching for systems that might have compact companions. 

Our analysis is focused on the short-period binaries, trying to avoid systems with primaries that are not on the main sequence (MS). This is because a large mass ratio is not necessarily an indication of a compact companion for giant stars, for example. 
As discussed by \citet*{gomel21b},
Algol-type binaries, with sub-giant or giant primaries, are famous counterexamples. These systems, which probably went through a mass-transfer phase during their evolution \citep[e.g.,][]{algol18, chen20}, can have a mass ratio larger than unity and still have a MS secondary \citep[e.g.,][]{negu18,samadi18}. 

%

In these binaries a giant, sub-giant or stripped giant primary is the less massive component, but nevertheless the brighter star of the system \citep[e.g.,][]{nelson01, budding04, mennekens17}.
Indeed, the recently suggested systems consisting of an evolved primary and a dormant compact-object secondary \citep[e.g.,][]{thompson19, liu19, rivinius20, jayasinghe21}, could be Algol-type binaries or even a binary with stripped giant companion \citep[e.g.,][]{van-den-Heuvel20, irrgang20, shenar20, bodensteiner20, mazeh20, el-badry20,elbadry22,elbadry22a}.  
Therefore, a more restricted list of candidates should include only stars that lie on or near the main sequence, with MMRs larger than unity. This implies to spectroscopic and photometric candidates alike.


The analysis resulted in a catalogue of short-period binary candidates that might have compact companions. 

Section~\ref{sec:choosing} lists the constraints used in order to define our catalogue of $6306$ candidate binaries that have mMMR larger than $0.5$ and therefore might have compact object companions, provided their primaries are on the MS and their modulation is induced by the ellipsoidal effect.
Section~\ref{sec:sample} describes the characteristics of the catalogue, Section~\ref{sec:CMD} examines the CMD location of a subsample of our candidates and Section ~\ref{sec:Simbad} cross-matches the catalogue with the available information found in \gaia DR2, Simbad, VSX catalogue and in the Chandra catalogue of X-ray sources. In Section~\ref{sec:best}, we narrow the catalogue and identify $262$ binaries with mMMR significantly larger than unity, and present their folded light curves and fitted ellipsoidal modulations. Section~\ref{sec:example} presents some details of three stars from the catalogue as examples, comparing the \gaia light curves with OGLE, ASAS-SN and ZTF photometry. We denote these stars with red throughout the diagrams of the paper. 
Finally, Section~\ref{sec:discussion} discusses and summarises our findings.

\section{Defining the catalogue}
\label{sec:choosing}

The compact-companion software, which is part of the variability-analysis software \citep[VariPipe, ][]{DR3-DPACP-162}, identified $\sim 20$ million possible ellipsoidal systems, obtained by the union of multiple classifiers \citep{DR3-DPACP-165}, with different ellipsoidal-probabilities selection cuts. We searched  this large sample for ellipsoidals with possible compact object companions. 

In our analysis we used mainly the \gmag light curve \citep{DR3-documentation}, as  only a small fraction of systems had enough \gbp or \grp measurements to derive significant amplitudes. 
For those, we could not see any colour-based clustering, and therefore used the colour information only to remove three outlier systems.

We used the cleaned  \gmag light curve, folded with $P$, twice the period found by a Generalised Lomb Scargle (GLS) period search \citep{zechmeister17}, as we assumed the \gaia pipeline detected the predominant ellipsoidal second-harmonic periodicity. The period uncertainty, $P_{\rm err}$, was derived by the pipeline.
%

We included only systems with:
\begin{enumerate}
    \item Number of cleaned \gmag field-of-view (FoV) transits > 25
    \item $0.25 < P < 2.5$ d
    \item $P / P_{\rm err} > 10$
	\item Frequencygram peak\footnote{measures the height of the frequencygram peak, relative to the frequencygram mean, in units of the standard deviation of the frequencygram.} > 12,
\end{enumerate}
%
to make sure the adopted period is secure and significant. The upper limit of condition 2 was used to avoid systems with non-MS primaries.
As shown by \citet{gomel21c}, the period of ellipsoidal variables with MS primaries is in most cases shorter than $2.5$ days. At this stage, we were left with $112\ 591$ candidates.

Next, a three-harmonic model was fitted to each light curve:
\begin{equation}
\gmag_{\rm mag} = \overline{G} + \ 
\sum_{i=1}^3 a_{i\mathrm{c}} \cos\big(\frac{2 \pi i}{P} (t-T_0)\big) + \ 
a_{i\mathrm{s}} \sin\big(\frac{2 \pi i}{P} (t-T_0)\big)\ ,
\label{eq:harmonics}
\end{equation}
with 7 free parameters, $\overline{G}, a_{i\mathrm{c}}, a_{i\mathrm{s}}$, $\{i=1,2,3\}$, that characterize the presumed periodic modulation. Each parameter was derived with its corresponding uncertainty.
$T_0$ was chosen so that $a_{2\mathrm{s}} = 0$. We defined the amplitude of each harmonic as
\begin{equation}
A_i = \ 
\sqrt{a^2_{i\mathrm{c}} + a^2_{i\mathrm{s}}} \ \ 
\{i = 1, 2, 3\} \ .
\label{eq:semiAmp}
\end{equation}

In the next stage of searching for candidates with compact-object companions we chose only  systems with:
\begin{enumerate}
    \item $0.33 < A_2 / (\rm range \ of \ G) < 0.6$
    \item $A_2 / A_{\rm 2,err} > 10$
    \item $A_1 / A_{\rm 1,err} > 3$ or $A_3 / A_{\rm 3,err} > 3$
	\item $A_1/A_2 < 1$ and $A_3/A_2 < 0.3$.
\end{enumerate}

The first two conditions were set to obtain a reliable and significant $A_2$. Condition 3 reflected our expectation for an ellipsoidal light curve to display non-equal minima, and the harmonic ratio of condition 4 is typical for ellipsoidals. We were left with a sample of $22\ 914$ systems that are highly probable short-period ellipsoidal variables.

\begin{figure} 
\centering
{\includegraphics[width=0.5\textwidth]{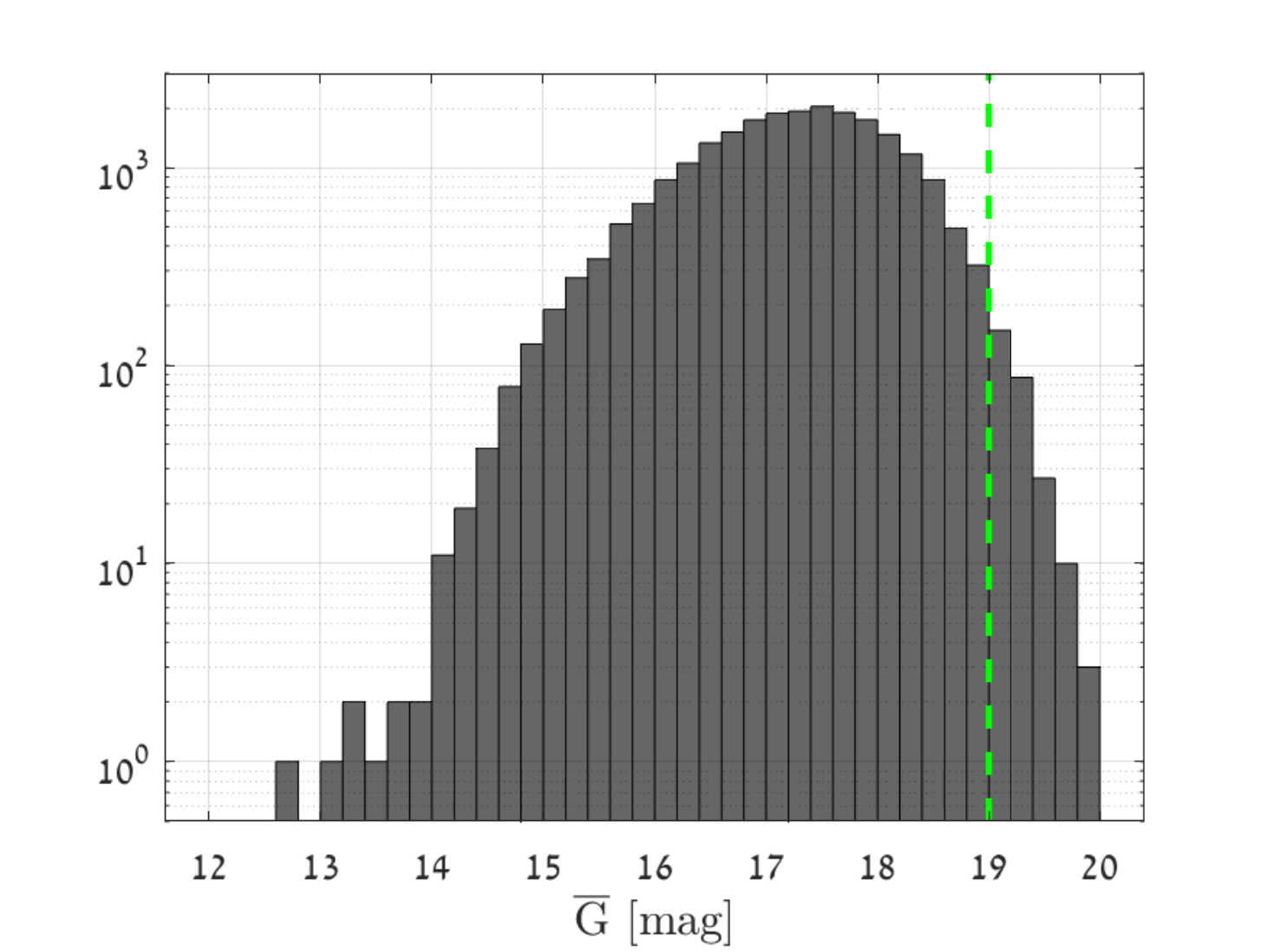}}
\caption{
Histogram of $\overline{G}$ of $22\ 914$ compact-companion candidates. 
The 19th mag  vertical green dashed line marks the observational limit of a few existing multi-object spectrographs that can be used to follow the candidates (see discussion in Section~\ref{sec:discussion}).
}
\label{fig:Gmag-23K-hist}
\end{figure}

The $\overline{G}$-mag histogram of the $22\ 914$ systems is shown in Fig.~\ref{fig:Gmag-23K-hist}. Note that this magnitude was derived by equation~(\ref{eq:harmonics}) and is slightly different from a simple arithmetic mean. It is evident that most systems are relatively faint, covering a $\overline{G}$-mag range of $13$--$20$.
In the figure we added the 19th mag mark, to denote the observational limit of a few existing  multi-object spectrographs that can be used to follow the best candidates (see discussion in Section~\ref{sec:discussion}). 

The final stage of constructing the catalogue of candidates was to identify ellipsoidal that might have compact companions. This was done by using equation~(1) of \citet*{gomel21b} that estimates the ellipsoidal leading amplitude $A_2$ as a function of the fillout factor $f$ --- the primary
volume-averaged radius divided by its Roche-lobe volume-averaged radius \citep{kopal59,paczynski71}, the orbital inclination $i$, and the secondary-to-primary mass ratio $q$: 
%
\begin{equation}
A_2 \simeq
\frac{1}{\overline{L}/L_{_0}}\alpha_\mathrm{2} \
f^3 E^3(q) \ q \ \sin^2 i \ C(q,f) \, ,
\label{eq:A2}
\end{equation}
%
where 
$\overline{L}$ is the average luminosity of the star, $L_{_0}$ is the stellar brightness with no secondary at all, and
$E(q)$ is the \cite{eggleton83} approximation for the volume-averaged Roche-lobe radius in binary semi-major axis units.
The ellipsoidal coefficient $\alpha_2$ depends on the linear limb- and gravity-darkening coefficients of the primary and is expected to be in the $1$--$2$ range. 
The correction coefficient $C(q,f)$ starts at $1$ for $f = 0$ (no correction), as expected, and rises monotonically as $f\to 1$, obtaining a value of $\sim 1.5$ at $f \gtrsim 0.9$ \citep*{gomel21a}.

Assuming a fillout factor of $f = 0.95$, inclination of $90^{\circ}$ and a typical $\alpha_2$ of 1.3 for the \gmag-band \citep{claret19}, we solved for the modified minimum mass ratio --- mMMR, $\hat{q}_{\rm min}$, \citep*{gomel21b},  
based on the observed second harmonic amplitude $A_2$.
We set $f$ to be $0.95$ to enable using the analytical approximation of \citet*{gomel21a}. Note that the mMMR value obtained for a fillout factor of unity would be smaller than the one we derived.

The  uncertainty of $\hat{q}_{\rm min}$ is inherently large because of the asymptotic increase of $A_2$ as a function of $q$, and due to the uncertainty in the ellipsoidal-amplitude approximation we use. In particular, $\alpha_2$ is not well known, so we adopted its uncertainty to be $0.1$. We then derived the uncertainty of $A_2/\alpha_2$, and assuming a Gaussian distribution for this ratio, using equation (\ref{eq:A2}),  we obtained $\hat{q}_{\rm min}^{-1\sigma}$ that corresponds to the $15.9$ percentile of $A_2/\alpha_2$. In a similar way we derived $\hat{q}_{\rm min}^{-3\sigma}$ that corresponds to the $0.135$ percentile of $A_2/\alpha_2$.


%

As pointed out by \citet*{gomel21b}, the mMMR is expected to be smaller than the actual mass ratio of the system, as it assumes a fillout factor close to unity and an inclination of $90^o$. Therefore, we opted, somewhat arbitrarily, to include systems with $\hat{q}_{\rm min} > 0.5$ in our catalogue of ellipsoidals that might have compact companions, resulting in $6336$ candidates. Obviously, 
different thresholds would have yielded different catalogues. In particular, adopting a different fillout factor, of, say, $f=0.98$, for solving 
equation (\ref{eq:A2}) would yield a smaller catalogue.

The adopted catalogue included $27$ sources that were identified as \rrl variables --- $25$ of them are known \rrl stars in the Milky-Way bulge, $24$ from the OGLE survey \citep{soszy14} and one from the VVV survey \citep{ramos18}. Two additional sources are not known in the literature but were confirmed as \rrl stars by inspecting their light curves. Therefore the $27$ presumably \rrl variables were excluded from the final list. 

Analysis of the \grp and \gbp light curves yielded only 
a small number of systems with periodic colour information significant enough for our use.
We derived the second-harmonic amplitude $A_{\rm 2, RP}$ by using equation (\ref{eq:harmonics}) for the \grp light curve, and inspected the $A_{\rm 2, RP}/A_{\rm 2}$ distribution for the $815$ systems with $A_{\rm 2, RP}/A_{\rm 2, RP, err}>10$. We found three outliers, with amplitude ratio smaller than $0.7$, that were excluded from the catalogue, leaving $6306$ candidates of ellipsoidal variables with compact-companion secondaries.

%

The resulting catalogue of $6306$ short-period ellipsoidals is given online in the DR3 archival data, in table \texttt{gaia\_dr3.vari\_compact\_companion}. The first $15$ candidates light curves and parameters are given in Fig.~\ref{fig:lcs} and Table~\ref{tab:Data}. 

\section{The catalogue}
\label{sec:sample}

This section presents some characteristics of the catalogue of the compact-companion candidates.

\begin{figure} 
\centering
{\includegraphics[width=0.5\textwidth]{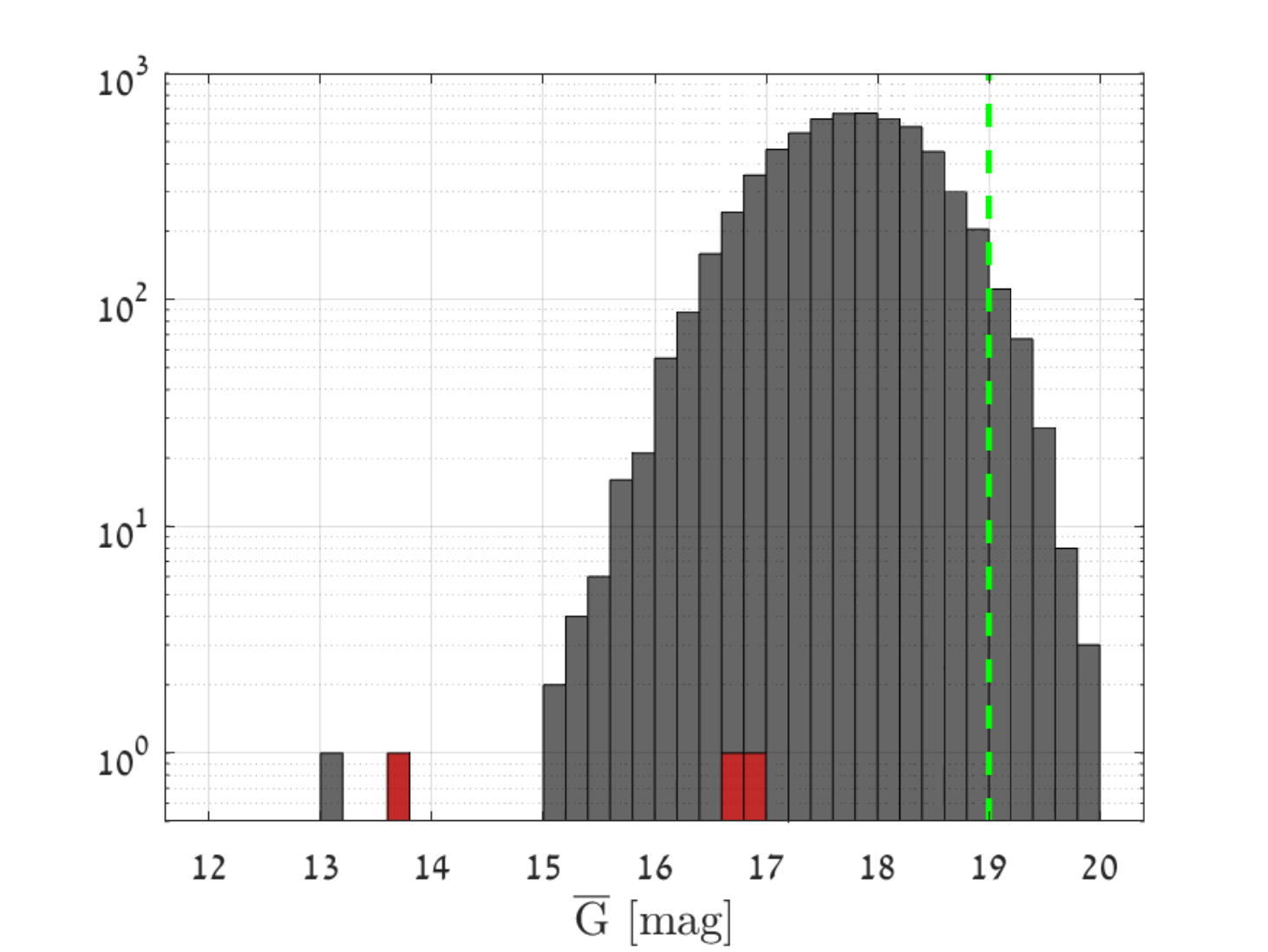}}
\caption{
Histogram of $\overline{G}$ of $6306$ compact-companion candidates. The three sources discussed in Section~\ref{sec:example} are colored in red. For the green line see Fig.~\ref{fig:Gmag-23K-hist}.
}
\label{fig:Gmag-hist}
\end{figure}

Fig.~\ref{fig:Gmag-hist} shows the $\overline{G}$-mag histogram of the candidates, as derived by equation~(\ref{eq:harmonics}).
The figure shows that most of the compact-companion candidates are relatively faint, in a $\overline{G}$-mag range of $15$--$20$, with two brighter sources within $13$--$14$ mag, one of which is discussed in Section~\ref{sec:example}.

\begin{figure} 
\centering
{  \includegraphics[width=0.5\textwidth]{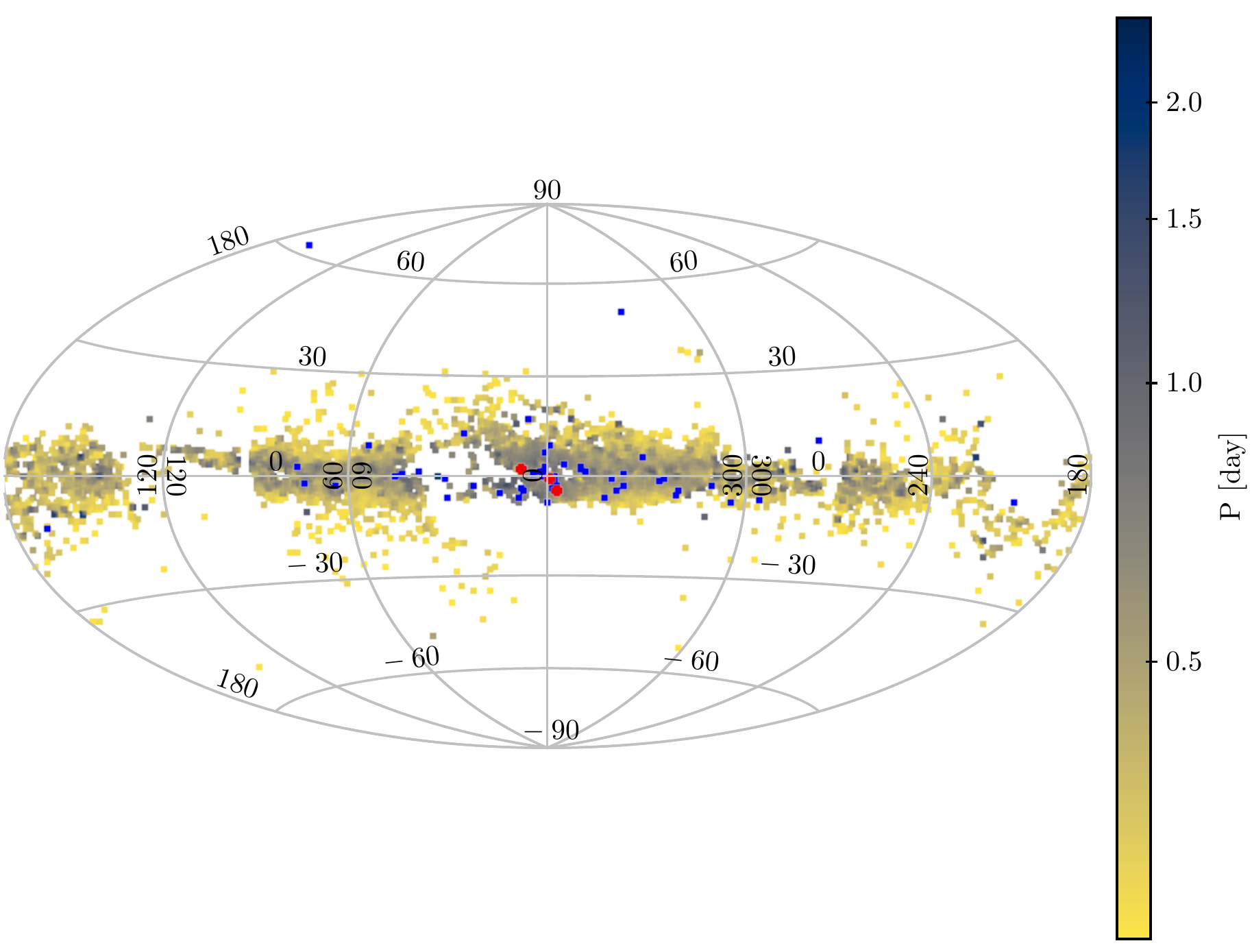}  }
\caption{Density distribution of the $6306$ compact-companion candidates in Galactic coordinates, colored by orbital period. The three sources discussed in Section~\ref{sec:example} are colored in red. The 59 Galactic X-ray binaries are shown by blue points.
The conspicuous circular hole between Galactic longitudes of $\sim 10$ and $\sim40$ degrees is due to \gaia scan law and our imposed lower limit of $25$ FoV transits. }
	\label{fig:SkyMap}
\end{figure}

A density distribution of the $6306$ catalog stars in Galactic coordinates, colored by the orbital period, is presented in Fig.~\ref{fig:SkyMap}.
One can see that the candidates are spread over the Galactic disk, most of them located towards the Bulge. Interestingly, the 59 known X-ray binaries \citep[e.g.,][]{corral16} are also concentrated in the Galactic disk, as seen in the figure.


\begin{figure} 
	\centering
	{  \includegraphics[width=0.5\textwidth]{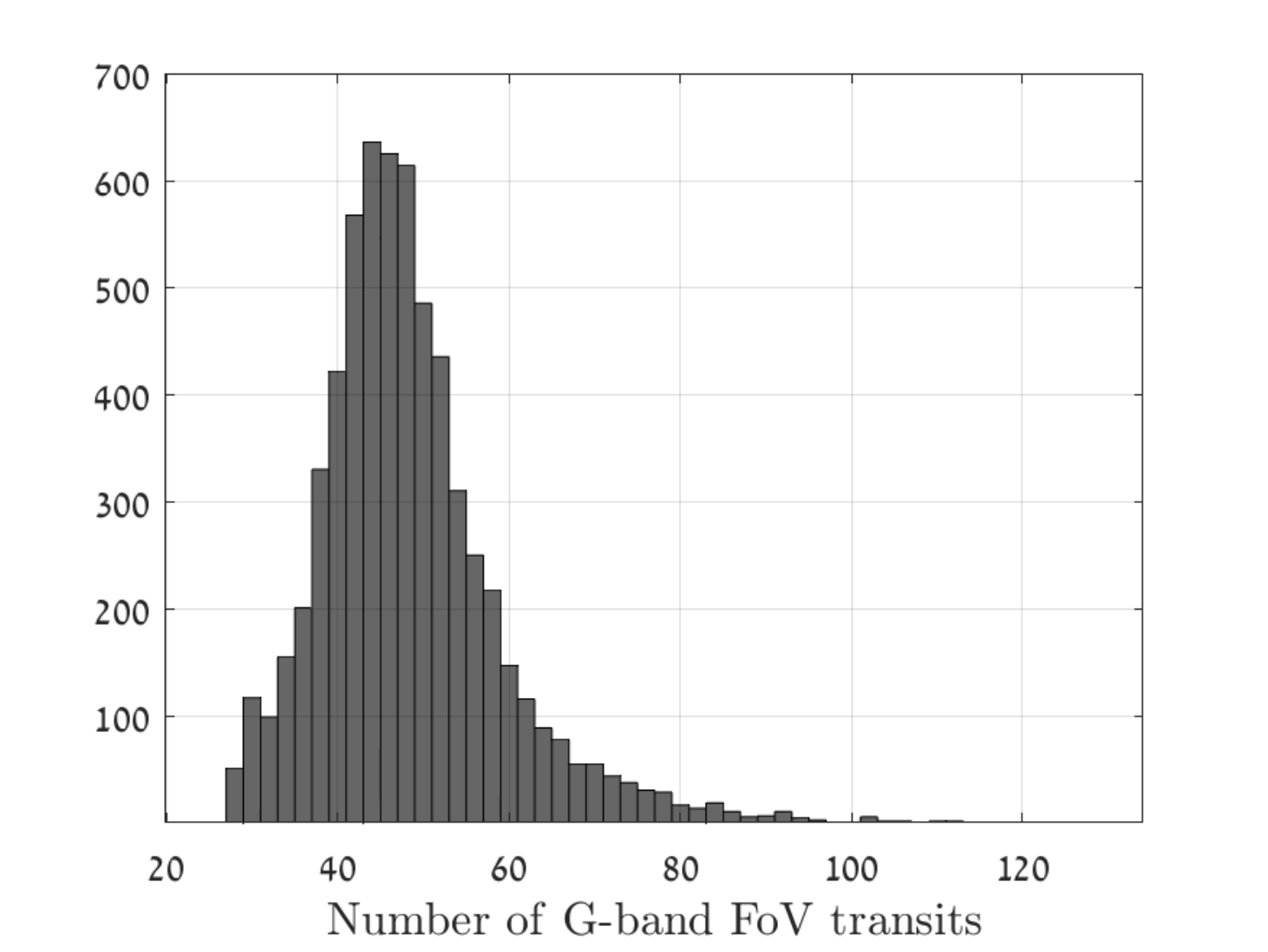}  }
	\caption{Histogram of number of \gmag-band FoV transits for $6306$ compact-companion candidates. The lower limit of $N=25$ was imposed by our analysis.}
	\label{fig:N-hist}
\end{figure}

Obviously, our ability to identify ellipsoidals, and compact-object candidates in particular, depends on the number of FoV transits. This is reflected in Fig.~\ref{fig:N-hist} that displays the distribution of number of \gmag-band FoV transits for the $6306$ candidates. Our imposed lower limit of $N=25$ is evident. The number of \gmag-band FoV transits varies between $25$ to $130$ with a median value of $46$. 

\begin{figure*} 
\centering
{\includegraphics[scale=0.8]{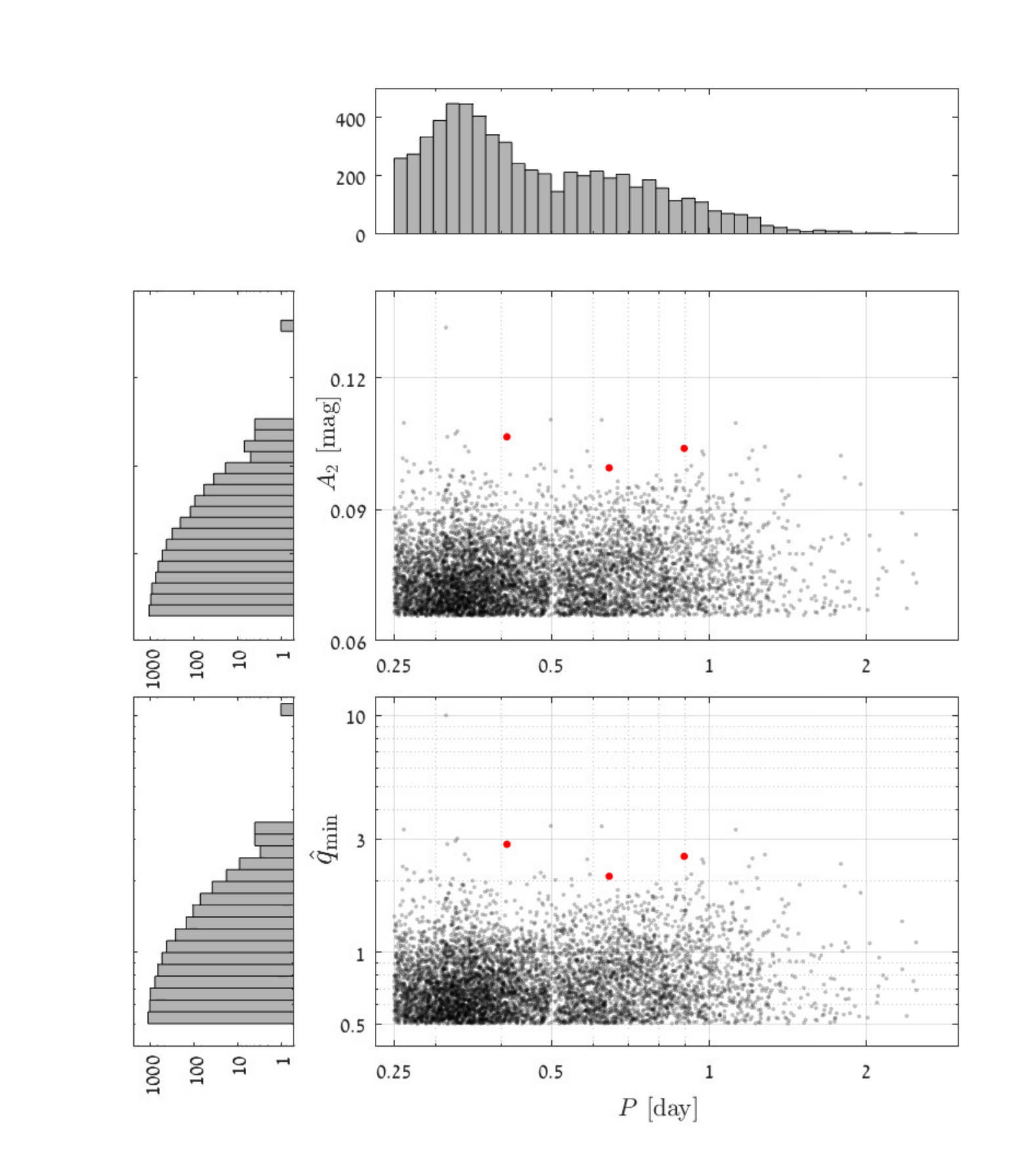}}
\caption{Second-harmonic amplitude ($A_2$) and mMMR ($\hat{q}_{min}$) as a function of orbital period ($P$) for the $6306$ compact-companion candidates. The mMMR was derived using a typical $\alpha_2$ of $1.3$ for the \gmag-band and assuming a fillout factor of 0.95. The three sources discussed in Section~\ref{sec:example} are colored in red.
	}
	\label{fig:P-A2-qmin}
\end{figure*}

The amplitude of the second harmonic $A_2$ and the mMMR, $\hat{q}_{min}$, are plotted in Fig.~\ref{fig:P-A2-qmin} as a function of the orbital period $P$. Each value of $A_2$ corresponds to a unique value of $\hat{q}_{min}$ as obtained from equation~(\ref{eq:A2}). The distributions of $A_2$, $\hat{q}_{min}$ and $P$ are plotted alongside their axes. 

As mentioned above, we focused on short-period systems with orbital periods between $0.25$--$2.5$ d, while the lower limit of $A_2$ $\sim 0.66$ mag corresponds to a minimum value of $\hat{q}_{min}$ = $0.5$, which defines our compact-companion candidate catalogue. 
Since we chose our catalogue to include only systems with relatively large $A_2$ and a significant frequencygram peak, we expect  a minor contamination, if at all, of spurious periods caused by \gaia scanning law.
A slight deficit of low-amplitude systems might be noticed in Fig.~\ref{fig:P-A2-qmin} around $P\sim 0.5$ d. This could be a result of our pipeline (Section~\ref{sec:choosing}) that was sensitive to the 
\gaia 6-hour orbital spin.


\section{Location of the candidates on the CMD}
\label{sec:CMD}
\begin{figure*} 
	\centering
	{\includegraphics[scale=0.9]{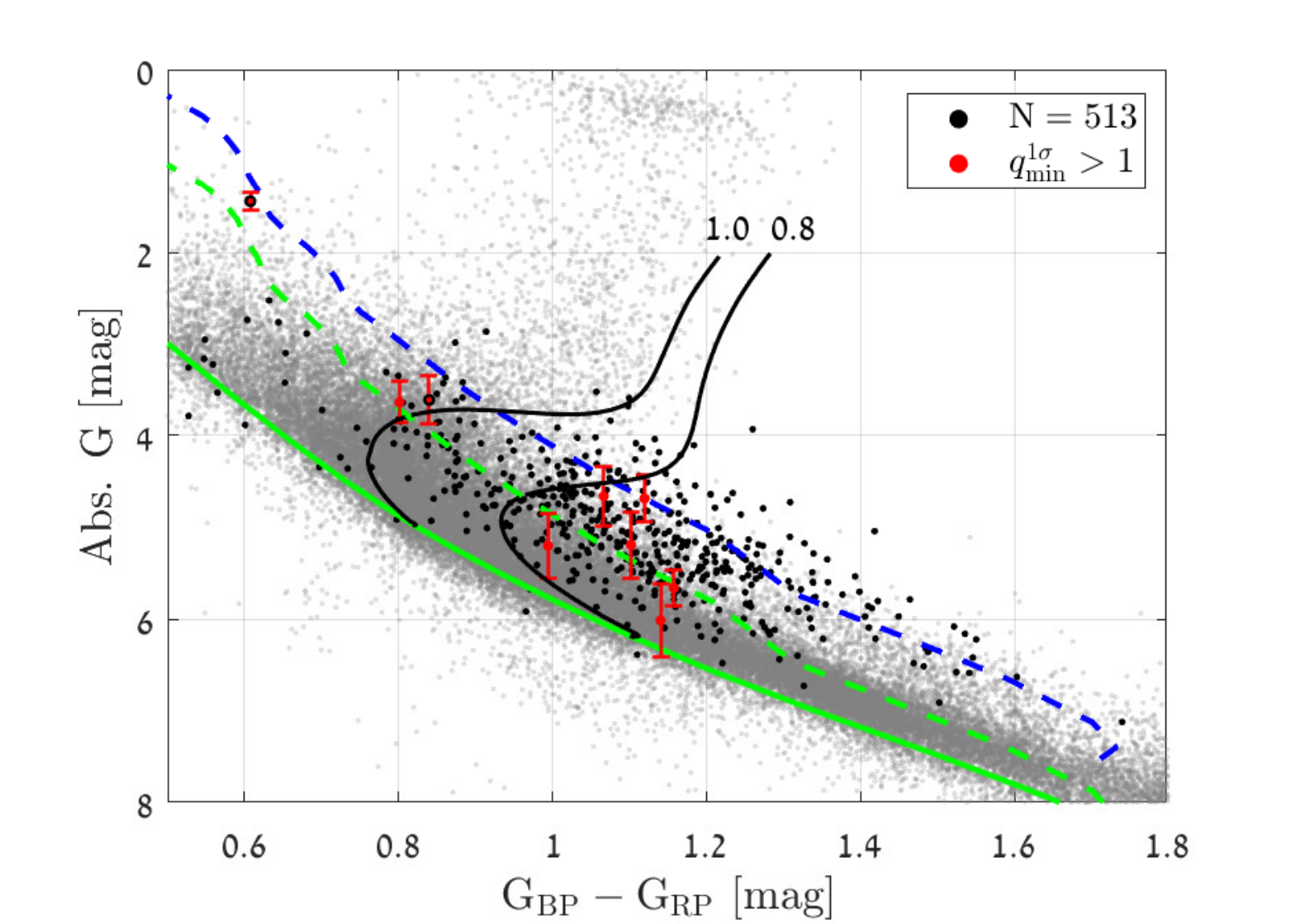}}
\caption{
Compact-companion candidates on the extinction-corrected
\gaia CMD. Only $513$ candidates with a minimum parallax of $0.5$ mas and parallax-over-error > 5 are plotted. Background grey-scale density map of stars with a minimum parallax of $0.5$ mas and parallax-over-error > 10 serves as a proxy for the expected CMD. 
Magnitudes and extinction corrections for the candidates and the background stars are from \gaia DR3 data \citep{Fouesneau22}.
Nine systems with  $\hat{q}_{\rm min} > 1$ are plotted in red (see text).
The two  green curves mark the theoretical expected edges of the MS.
The dashed-blue curve is the theoretical upper \gmag value expected for binaries with two identical MS stars, obtained by shifting the upper bound of the MS by 0.75 mag.
In black we marked the evolutionary tracks of $0.8$ and $1\, M_{\odot}$ stars within their MS phase, obtained by the \textsf{ISOCHRONE}\protect\footnote{https://github.com/timothydmorton/isochrones} software.
Two sources --- DR3 $4042390512917208960$
(upper point) and DR3 $4056017172771375616$ (lower point) discussed in Section~\ref{sec:example}, are colored in red with a black edge color.
%
%
%
}
\label{fig:CMD}
\end{figure*}

As mentioned above, we are interested in ellipsoidal variables with primaries that are not giants or sub-giants. To check if indeed this is the case we plotted in Fig.~\ref{fig:CMD} the position of a sub-sample of our candidates on the \gaia Colour Magnitude Diagram (CMD). Only $513$ candidates were included, for which the relative parallax precision is better than $20$\% (\parallax$/$ \sigparallax > $5$), the parallax is larger than $0.5$ mas (\parallax > 0.5 mas), and a \gaia extinction estimate is available.

The location of the sub-sample on the CMD is shown, after correcting for their extinction, taken from \gaia DR3 data \citep{Fouesneau22}. As a background we plotted the \gaia stars that have the same characteristics as the ellipsoidal sub-sample, but with a relative parallax precision better than $10$\%. The background stars mark the location of the MS, and even the red clump can be seen in the upper part of the figure. We also added two theoretical (green) curves that mark the expected edges of the MS, as derived with the \textsf{ISOCHRONE}\footnote{https://github.com/timothydmorton/isochrones}
software package \citep{morton15}. 

Fig.~\ref{fig:CMD} clearly shows that, at least for the sub-sample plotted, we succeeded to avoid the giant and the sub-giant branch of the CMD, probably due to the period limit of $2.5$ days. Therefore, {\it if} the candidates are indeed ellipsoidals and {\it if} their mass ratios are larger than unity, they {\it might} have compact companions. 

Note, however, that the minimum mass ratio of our sample is $0.5$, as we chose candidates with $\hat{q}_{\rm min} > 0.5$, and therefore the secondary of many of our candidates can still be a MS component. This is probably reflected in the figure, where the location of quite a few of the candidates is {\it above} the MS, possibly due to the light contribution of the MS secondary. To show this point we plotted in the figure (dashed-blue curve) the theoretical upper \gmag value expected for binaries with two MS stars, which was obtained by shifting the upper bound of the MS by 0.75 mag.

However, we emphasize that CMD locations can suffer from large uncertainties in absolute \gmag, due to errors in parallax, photometry and extinction estimates that may be highly inaccurate \citep{GaiaDR2,andrae18,anders19}. In addition, the extinction uncertainty can shift the \gbp $-$ \grp colour index. 
We demonstrate this point by plotting nine candidates for which the minimum mass ratio is larger than unity, as discussed in Section~\ref{sec:best}. Their companions are less likely to be on the MS, and nevertheless the locations of six of them  appear above the MS. For those, we plot the \gmag $1\sigma$ uncertainty, showing that all of the six systems might still be, within $1$ or $2\sigma$, on the MS strip. 
 
We note that the figure suggests that many of the candidates in our sample might be with mass, say, smaller than $0.8\, M_{\odot}$.  Consequently,  their compact companions might be with relatively small mass, in the white-dwarf range.

\section{Cross-match with \gaia DR2, Simbad, Chandra and VSX}
\label{sec:Simbad}

In this section we cross-match our catalogue with the Simbad astronomical database,\footnote{https://simbad.u-strasbg.fr/simbad/} \gaia DR2 variability data set \citep{GaiaDR2}, Chandra source Catalog,\footnote{https://cxc.cfa.harvard.edu/csc/} release 2.0, and the International Variable Star Index (VSX) catalogue.\footnote{https://www.aavso.org/vsx/}

\subsection{Cross-match with the Simbad catalogue}

The cross-match with Simbad, using a search radius of $3\arcsec$,
resulted in $528$ systems, summarized in Table~\ref{tab:SimVar}.  The cross-match yielded $522$ stars classified as variables, and six additional systems with no main identifier of variability. Two out of the six systems were detected by the Chandra mission \citep{evans10} as having significant X-ray emission.

\renewcommand{\arraystretch}{1.5}
\begin{table*}
\caption{Cross-match classifications of $528$ candidates obtained from Simbad database 
\label{tab:SimVar}}
\centering 
\begin{tabular}{l r l} 
\hline
Classification & Number & Source \\ 
\hline 
Eclipsing binaries      & $431$ & OGLE survey \\ 
                        & $9$   & CRTS survey \citep{drake14} \\
                        & $7$   & BEST II catalogue of variable stars \citep{fruth13} \\
                        & $2$   & \citet{miller10} \\
                        & $1$   & General catalogue of variable stars \citep{samus17} \\
                        & $1$   & \citet{deMarchi10} \\
                        & $1$   & \citet{weldrake08} \\
\hline
Ellipsoidal variables   & $65$  & OGLE survey \\
\hline
\rrl stars          & $2$  & OGLE survey \\
\hline
Rotational variable     & $1$   & CRTS survey \citep{drake14} \\
\hline
Delta Scuti             & $1$   & BEST II catalogue of variable stars \citep{fruth13} \\
\hline
Orion-type variable     & $1$   & \citet{rebull20} \\
\hline
X-ray sources           & $2$   & \citet{muno03,muno06} \\
\hline 
Other stars             & $4$   & \citet{metzger98,bernabei01}; \\
                        &       & \citet{audard07,kuhn17} \\
\hline 
\end{tabular}
\end{table*}

Most systems common to our and Simbad catalogues were classified by the OGLE team as eclipsing binaries. We inspected a few tens of them and found in all cases that 
the OGLE $I$ and \gaia \gmag light curves did not present clear narrow or flat-bottom eclipses. They showed possible, if at all, very broad two V-shape eclipses, centered around phases 0 and 0.5, suggesting they might be either contact binaries or ellipsoidal variables. It is not easy to differentiate between the two modulations, as both are spread over the entire binary-period phase, and have similar shapes.
Nevertheless, we suggest that the contamination of the catalogue by contact binaries is small, as discussed in Section~\ref{sec:discussion}.

We visually inspected the modulation of the two stars identified by OGLE as \rrl variables (\gaia DR3 $4116610292178919936$, $4107297257038868736$), and found that the OGLE classification is more probable, at least for the second object. The fact that only two cases were found as \rrl stars out of $528$ cross-matched systems suggests that  
about half a percent of the stars of our catalogue might be \rrl variables.

\subsection{Cross-match with \gaia DR2 variables} 
\label{sec:GaiaDR2var}

A cross-match of the candidate list with \gaia DR2 variability data set \citep{GaiaDR2}, resulted in 2 systems,  $2024410711652386560$ and $241721038597143296$. 
The two sources were classified as rotational variables, with a rotational period that is half the value presented in our catalogue,
 but our analysis suggests they are really ellipsoidal variables. None of the sources appears in the Simbad database.

\subsection{Cross-match with Chandra Source Catalogue}
\label{sec:Xray}

A cross-match of the candidate list with the Chandra Source Catalogue \citep{evans10}, using a search radius of $1\arcsec$, resulted in 3 systems, presented in Table~\ref{tab:CSC}, all included in Table~\ref{tab:SimVar}. The third object, \gaia DR3 $5966509571940818048$, is in the "Other stars" category, classified by Simbad as a young stellar-object candidate. Its X-ray flux is $\sim 2\times 10^{-15}$ erg/cm$^2$/s, with a relatively soft spectrum, consistent with the X-ray luminosity (derived from its \gaia EDR3 parallax) coming from the stellar surface. The other two X-ray sources are too faint for a significant flux measurement.  

\renewcommand{\arraystretch}{1.5}
\begin{table}
\caption{Cross-match with Chandra Source Catalogue}
\label{tab:CSC}
\centering 
\begin{tabular}{l l} 
\hline
\gaia DR3 & Name \\ 
\hline 
$4056853999874300544$ & 2CXO J174344.8-295445 \\ \hline 
$4057484501078108800$ & 2CXO J174512.0-285756 \\ \hline
$5966509571940818048$ & 2CXO J165419.4-414805 \\ \hline
\end{tabular}
\end{table}

\subsection{Cross-match with the VSX catalogue}
\label{sec:VSX}

\begin{figure*} 
\centering
{\includegraphics[scale=0.6]{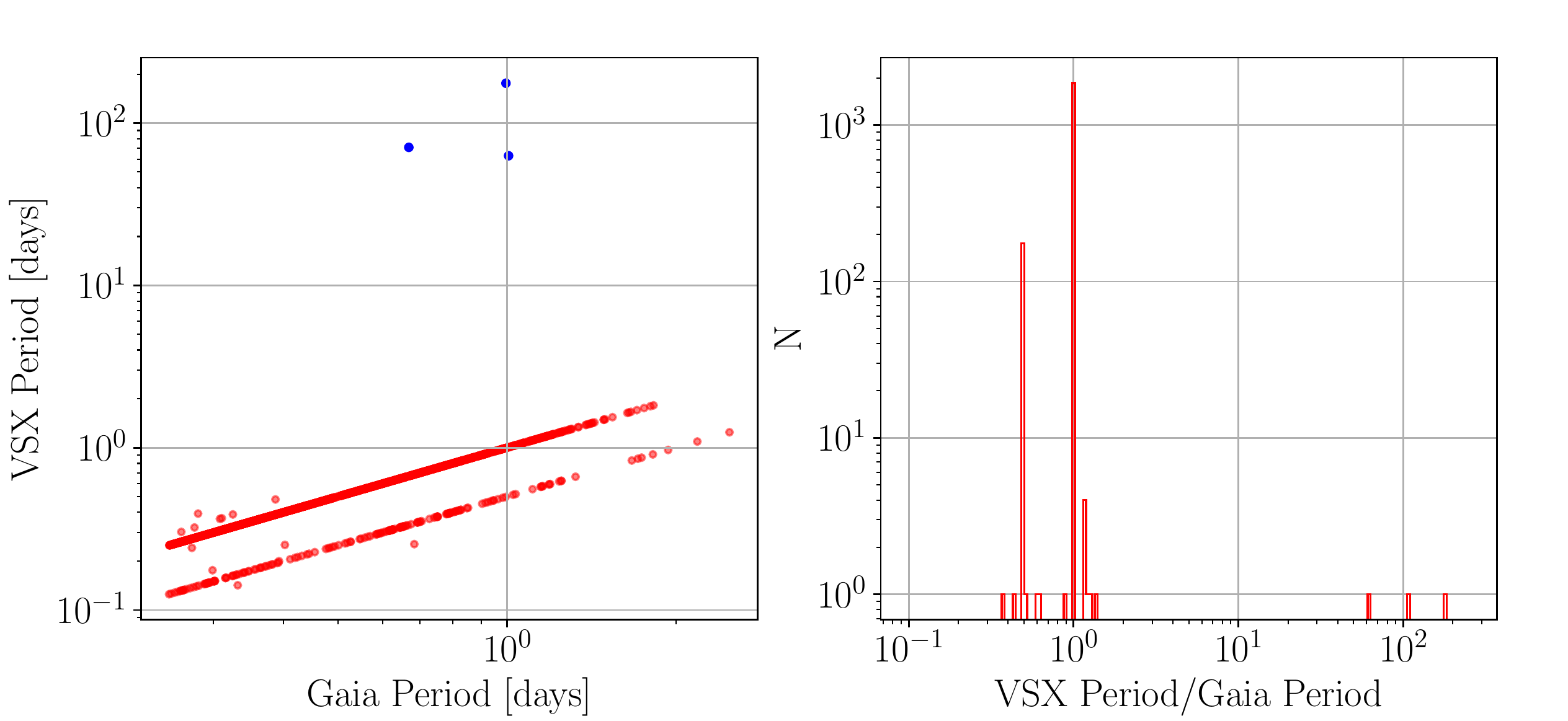}}
\caption{Scatter plot (left) and histogram (right) of the $2044$ VSX periods vs.~the Gaia ones.}

	\label{fig:VSX_Gaia_per}
\end{figure*}

Following the referee's suggestion, we cross-matched our candidates with the International Variable Star Index (VSX) catalogue,\footnote{https://www.aavso.org/vsx/} and found $2044$ common systems with derived periodicities and available light curves. Figure \ref{fig:VSX_Gaia_per} shows the VSX periods vs.~the Gaia ones. As can be seen, most systems ($1853$) have identical periods in the two catalogues. Some systems ($175$) display a VSX period which is half the \gaia period, consistent with them being ellipsoidal variables for which the dominant frequency of the modulation is the second harmonic.

\begin{figure*} 
\centering
{\includegraphics[scale=0.7]{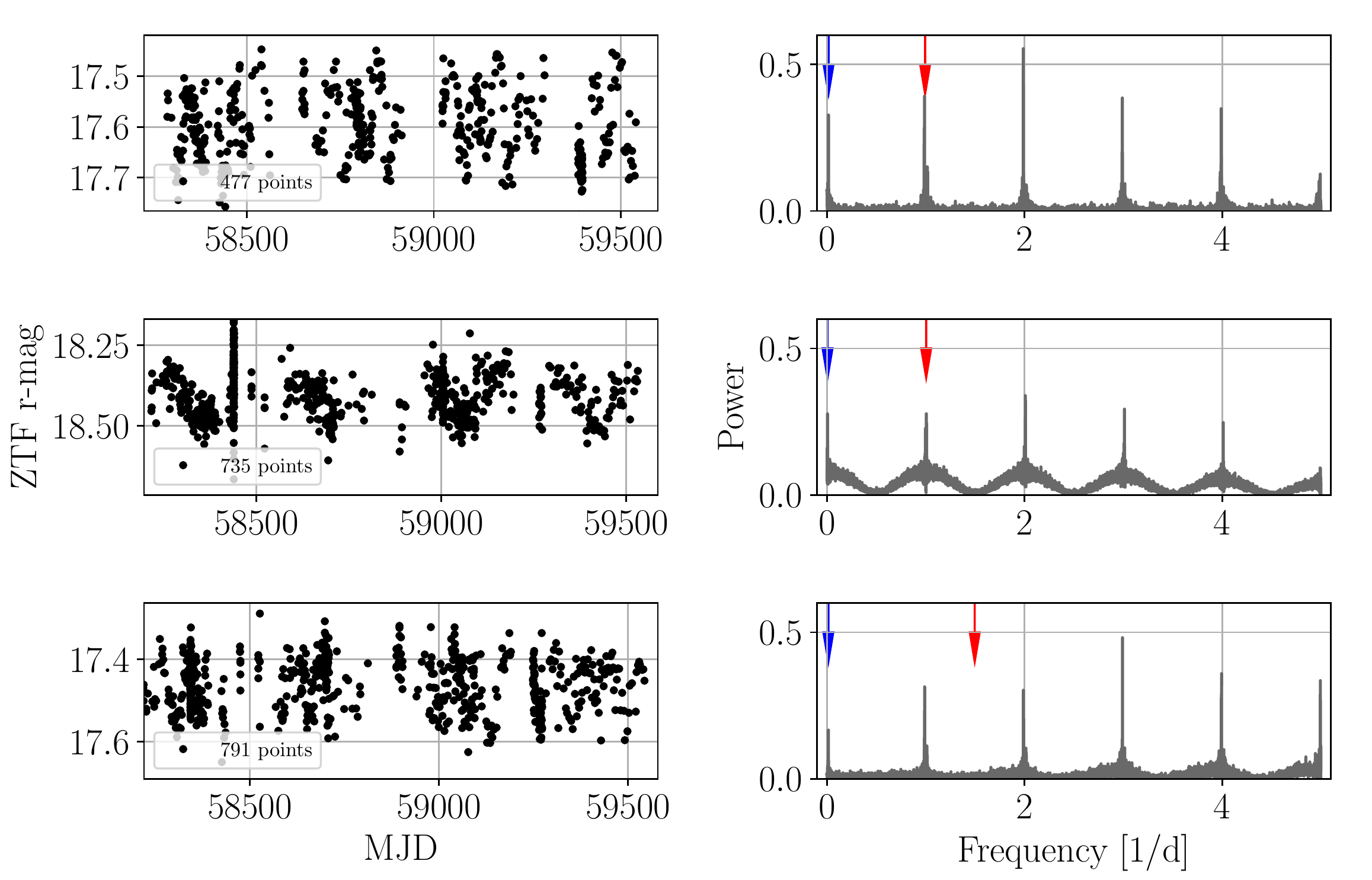}}
\caption{Time series (left) and power spectra (right) of the ZTF project data for the three outlier sources. 
The red (blue) vertical line marks the frequency corresponding to the  derived \gaia (ZFT) period.}
\label{fig:VSX_Gaia_outliers}
\end{figure*}

Three sources (marked in blue circles of Figure \ref{fig:VSX_Gaia_per}), \gaia DR3 $461521311430643456$, \gaia DR3 $1834964858132757504$ and \gaia DR3 $1825053924741093376$, analysed by the ZTF project \citep{Bellm19}, display clear discrepancy between the \gaia and the VSX periods. 
We re-analyzed the ZTF light curves of the three systems\footnote{https://irsa.ipac.caltech.edu/Missions/ztf.html} by deriving their corresponding power spectra, as shown in Figure \ref{fig:VSX_Gaia_outliers}. The frequencies of the strongest peaks of the three ZTF power-spectra
are within $1\sigma$ of the corresponding \gaia ones.

The spectra are dominated by strong side lobes on the two sides of the strongest peak, separated by $\Delta f=1$, reflecting the daily window of ground-based observations.
The frequencies of the highest peaks of the three systems, which correspond to half the \gaia orbital periods, are close to a whole number, in units of $1/day$, and therefore have one side-lobe peak with a low frequency, as seen in the figure. That side lobe corresponds to the period suggested by the ZTF analysis.
This exercise shows the potential advantage of space-mission measurements, that are not subject to the daily cycle of ground-based observations.

\section{Candidates with mMMR larger than unity}
\label{sec:best}

We chose $262$ ellipsoidals with $\hat{q}_{\rm min}^{-1\sigma}>1$, which we considered as having mMMR significantly larger than unity, and therefore are more promising candidates for having compact companions. 
Forty one of these candidates were found in Simbad --- 39 are classified as eclipsing binaries and 2 as ellipsoidals by the OGLE project \citep{soszy16}. One of the two ellipsoidals is relatively bright, with a \gmag magnitude of $\sim$ $13.8$, and is discussed in detail in Section~\ref{sec:example}.

\begin{figure*} 
\centering
{  \includegraphics[scale=0.8, trim={0 2cm 0 0}, clip]{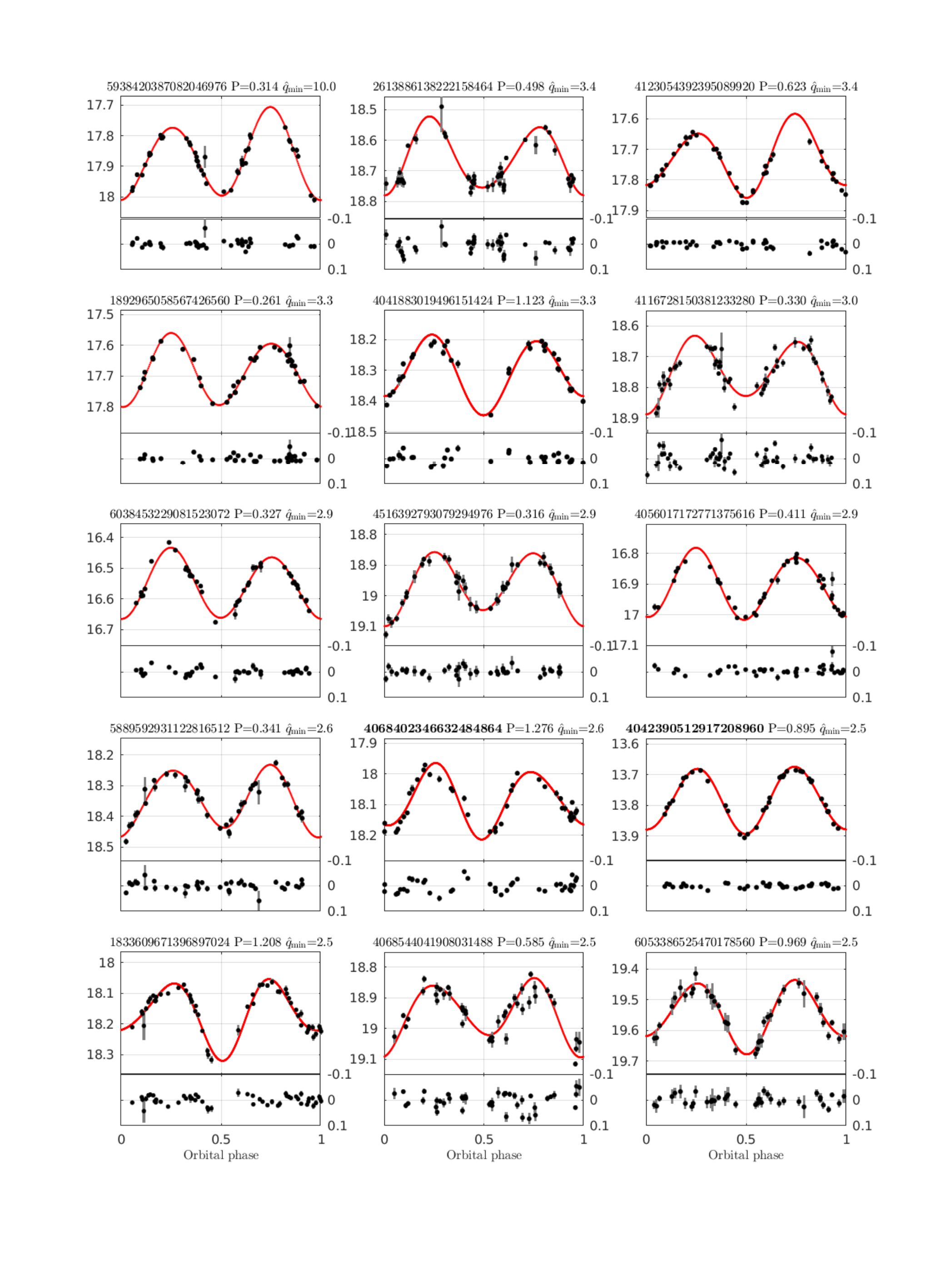}  }
\caption{ Folded \gaia light curves in the \gmag band of the first 15 candidates with $\hat{q}_{\rm min}^{-1\sigma}$ > $1$, in descending $\hat{q}_{\rm min}$ order. For each candidate, the \gaia DR3 id, together with the \gaia period in days and the value of $\hat{q}_{\rm min}$, are given. The epoch of the second-harmonic minimum corresponds to phases $0$ and $0.5$, and a three-harmonics model is plotted with a solid red line. All curves are plotted with a mag range of 0.4, for convenience. The residuals are plotted in the lower panels and are shown between -0.1 to 0.1 mag for clarity. Marked identifiers are of \gaia DR3 4068402346632484864, which might be an \rrl variable, discussed in Section~\ref{sec:best}, and \gaia DR3 4042390512917208960, which is one of our brightest candidates, discussed in Section~\ref{sec:example}.
}
\label{fig:lcs}
\end{figure*}

Folded light curves and a summary table for the first $15$ candidates, in descending $\hat{q}_{\rm min}$ order, are presented in Fig.~\ref{fig:lcs} and Table~\ref{tab:Data}. Similar figures and a table for all $262$ candidates are given in the online supplementary document. The table includes the \gaia DR3 identifier (id), \gaia orbital period, reference time $T_0$ [BJD-2455197.5] which was chosen so that $a_{2\mathrm{s}} = 0$, 
average G magnitude $\overline{G}$, 
and cosine and sine Fourier coefficients $a_{i\mathrm{c}}, a_{i\mathrm{s}}$ $\{i=1,2,3\}$ of the three-harmonic model, defined by equation~(\ref{eq:harmonics}), 
each with its uncertainty. The table also gives the total number of \gmag-band FoV transits, $N$, the derived mMMR and the lower-percentile mMMR.

All included parameters of Table~\ref{tab:Data}, except $N$, are given online in the DR3 archival data, in table \texttt{gaia\_dr3.vari\_compact\_companion}. The archival data includes also reference times, averaged magnitudes, and Fourier coefficients with their uncertainties for the \gaia \gbp and \grp light curves, the $0.135$th percentile mMMR, $\hat{q}_{\rm min}^{-3\sigma}$, and the ellipsoidal coefficient $\alpha_2$, defined in equation~(\ref{eq:A2}).

Out of the $262$ light curves, that of \gaia DR3 4068402346632484864, has an untypical shape for ellipsoidal variables, as seen in Fig.~\ref{fig:lcs} (marked in bold), suggesting this star is an \rrl variable. This is consistent with the above discussion that our catalogue is contaminated by $\sim 1$\% of \rrl stars. 

\newcommand*{\thead}[1]{\multicolumn{1}{|c|}{\bfseries #1}}
\renewcommand{\arraystretch}{2}
\begin{table*}[ht!]
\caption{ Fitted parameters of the first 15 candidates in descending $\hat{q}_{\rm min}$ order
	}
    \resizebox{\textwidth}{!}{
	\begin{tabular}{|r|r|r|r|r|r|r|r|r|r|r|r|r|}
		\hline
		 \thead{\gaia $\rm DR3$} & \thead{$P$} & \thead{$T_0$} & \thead{$\overline{G}$} & \thead{$a_{\rm 1c}$} & \thead{$a_{\rm 2c}$} & \thead{$a_{\rm 3c}$} & \thead{$a_{\rm 1s}$} & \thead{$a_{\rm 2s}$} & \thead{$a_{\rm 3s}$} & \thead{$N$} & \thead{$\hat{q}_{\rm min}$} & \thead{$\hat{q}_{\rm min}^{-1\sigma}$} \\
		 \thead{} & \thead{$P_{\rm err}$} & \thead{$T_{\rm 0,err}$} & \thead{$\overline{G}_{\rm err}$} & \thead{$a_{\rm 1c,err}$} & \thead{$a_{\rm 2c,err}$} & \thead{$a_{\rm 3c,err}$} & \thead{$a_{\rm 1s,err}$} & \thead{$a_{\rm 2s,err}$} & \thead{$a_{\rm 3s,err}$} & \thead{} & \thead{} & \thead{} \\
		 \thead{} & \thead{$\rm [day]$} & \thead{$\rm [BJD-2455197.5]$} & \thead{$\rm [mag]$} & \thead{$\rm [mag]$} & \thead{$\rm [mag]$} & \thead{$\rm [mag]$} & \thead{$\rm [mag]$} & \thead{$\rm [mag]$} & \thead{$\rm [mag]$} & \thead{} & \thead{} & \thead{} \\
		\hline
        5938420387082046976 & 0.314117 & 2244.03610 & 17.8719 & 0.0098 & 0.1315 & -0.0026 & 0.0226 & 0.0000 & -0.0115 & 43 & 10.0 & 5.6\\
        & 0.000054 & 0.00056 & 0.0018 & 0.0028 & 0.0036 & 0.0026 & 0.0033 & 0.0021 & 0.0035 &  &  & \\
        \hline 
        2613886138222158464 & 0.49774 & 2423.3476 & 18.6571 & -0.0103 & 0.1105 & 0.0229 & -0.0110 & 0.0000 & 0.0079 & 46 & 3.4 & 2.1\\
         & 0.00012 & 0.0020 & 0.0039 & 0.0047 & 0.0065 & 0.0067 & 0.0067 & 0.0046 & 0.0057 &  &  & \\
        \hline 
        4123054392395089920 & 0.62299 & 2425.4947 & 17.7272 & -0.0101 & 0.1104 & -0.0109 & 0.0220 & 0.0000 & -0.0105 & 43 & 3.4 & 2.2\\
         & 0.00021 & 0.0013 & 0.0031 & 0.0023 & 0.0047 & 0.0029 & 0.0055 & 0.0024 & 0.0038 &  &  & \\
        \hline 
        1892965058567426560 & 0.261008 & 2191.02880 & 17.6875 & 0.0023 & 0.1097 & 0.0008 & -0.0062 & 0.0000 & 0.0113 & 36 & 3.3 & 2.2\\
         & 0.000044 & 0.00049 & 0.0022 & 0.0033 & 0.0035 & 0.0028 & 0.0028 & 0.0026 & 0.0033 &  &  & \\
        \hline 
        4041883019496151424 & 1.12284 & 2355.7487 & 18.3060 & -0.0314 & 0.1097 & 0.0003 & -0.0068 & 0.0000 & 0.0041 & 34 & 3.3 & 2.1\\
         & 0.00077 & 0.0054 & 0.0036 & 0.0057 & 0.0061 & 0.0052 & 0.0055 & 0.0063 & 0.0054 &  &  & \\
        \hline 
        4116728150381233280 & 0.330273 & 2487.1291 & 18.7504 & 0.0175 & 0.1078 & 0.0123 & -0.0068 & 0.0000 & 0.0032 & 53 & 3.0 & 1.9\\
         & 0.000043 & 0.0014 & 0.0035 & 0.0059 & 0.0074 & 0.0061 & 0.0061 & 0.0040 & 0.0059 &  &  & \\
        \hline 
        6038453229081523072 & 0.327324 & 2105.02175 & 16.5562 & -0.0002 & 0.1073 & 0.0019 & -0.0101 & 0.0000 & 0.0057 & 41 & 2.9 & 2.0\\
         & 0.000082 & 0.00072 & 0.0020 & 0.0035 & 0.0042 & 0.0037 & 0.0033 & 0.0023 & 0.0031 &  &  & \\
        \hline 
        4516392793079294976 & 0.316059 & 2159.86061 & 18.9664 & 0.0214 & 0.1066 & 0.0048 & 0.0014 & 0.0000 & 0.0034 & 41 & 2.9 & 1.9\\
         & 0.000046 & 0.00074 & 0.0019 & 0.0032 & 0.0035 & 0.0029 & 0.0028 & 0.0026 & 0.0033 &  &  & \\
        \hline 
        4056017172771375616 & 0.410707 & 2394.25673 & 16.9046 & -0.0038 & 0.1066 & -0.0008 & -0.0049 & 0.0000 & 0.0112 & 44 & 2.9 & 1.9\\
         & 0.000093 & 0.00078 & 0.0016 & 0.0023 & 0.0031 & 0.0024 & 0.0027 & 0.0024 & 0.0028 &  &  & \\
        \hline 
        5889592931122816512 & 0.341463 & 2204.82654 & 18.3464 & 0.0162 & 0.1044 & -0.0004 & -0.0033 & 0.0000 & -0.0129 & 39 & 2.6 & 1.8\\
         & 0.000081 & 0.00084 & 0.0021 & 0.0031 & 0.0037 & 0.0030 & 0.0033 & 0.0026 & 0.0034 &  &  & \\
        \hline 
        4068402346632484864$^1$ & 1.27615 & 2356.9575 & 18.0847 & -0.0103 & 0.1043 & -0.0132 & -0.0015 & 0.0000 & 0.0135 & 39 & 2.6 & 1.6\\
         & 0.00079 & 0.0056 & 0.0044 & 0.0062 & 0.0070 & 0.0062 & 0.0064 & 0.0055 & 0.0061 &  &  & \\
        \hline 
        4042390512917208960$^2$ & 0.89522 & 2383.8996 & 13.7822 & 0.0000 & 0.1039 & -0.0074 & 0.0046 & 0.0000 & 0.0012 & 38 & 2.5 & 1.8\\
         & 0.00046 & 0.0014 & 0.0015 & 0.0022 & 0.0022 & 0.0023 & 0.0020 & 0.0020 & 0.0019 &  &  & \\
        \hline 
        1833609671396897024 & 1.20849 & 2202.2576 & 18.1657 & -0.0286 & 0.1035 & -0.0215 & -0.0010 & 0.0000 & -0.0083 & 53 & 2.5 & 1.7\\
         & 0.00074 & 0.0028 & 0.0028 & 0.0054 & 0.0048 & 0.0038 & 0.0030 & 0.0031 & 0.0038 &  &  & \\
        \hline 
        4068544041908031488 & 0.58456 & 2430.2668 & 18.9518 & 0.0258 & 0.1033 & 0.0116 & -0.0027 & 0.0000 & -0.0153 & 43 & 2.5 & 1.6\\
         & 0.00025 & 0.0026 & 0.0043 & 0.0063 & 0.0064 & 0.0064 & 0.0058 & 0.0057 & 0.0058 &  &  & \\
        \hline 
        6053386525470178560 & 0.96892 & 2206.1084 & 19.5447 & -0.0177 & 0.1032 & -0.0122 & 0.0016 & 0.0000 & -0.0040 & 40 & 2.5 & 1.6\\
         & 0.00052 & 0.0033 & 0.0035 & 0.0045 & 0.0055 & 0.0053 & 0.0053 & 0.0042 & 0.0047 &  &  & \\
        \hline  
	\end{tabular}
	}
	\\
	\vspace{3mm}
	\label{tab:Data}
	\begin{tablenotes}
      \tiny
      \item $^1$ Might be an \rrl variable, discussed in Section~\ref{sec:best}.
      \item $^2$ One of our brightest candidates, discussed in Section~\ref{sec:example}.
    \end{tablenotes}
\end{table*} 

A close look at Fig.~\ref{fig:lcs} might give the impression that the two maxima of some of the light curves have different heights. This is reflected by the difference between the coefficient of  $a_{\rm 1s}$ and $a_{\rm 3s}$, indicating either that the light curves are not due to an ellipsoidal modulation, or that we are witnessing an additional effect. However, this is probably not the case. The \gaia data is quite sparse and in many cases not enough to identify such local small features of the modulation. One good example is  \gaia DR3 4056017172771375616, which is discussed in the next section. One can see in Fig.~\ref{fig:LC_5616} that the \gaia model, based on 44 measurements only, does display two different peaks, while the OGLE light curve, which has 70 points, shows equal maxima.

\section{Three examples of compact-companion candidates}
\label{sec:example}

This section concentrates on three candidates from the catalogue, examining their photometric modulations as obtained by other surveys --- OGLE \citep{soszy16, pawlak16}, ASAS-SN \citep{Shappee14,kochanek17} and ZTF \citep{Bellm19}, and compare them with that of \gaia. Then the reliability of their candidacy as having compact-object companions is discussed, exemplifying the potential of the candidates in the catalogue. 

The three stars are marked in the pertinent figures above, except for \gaia DR3 4070409432055253760, 
that does not have an estimation of its extinction and therefore does not appear in Figure~\ref{fig:CMD}. 
\gaia DR3 4042390512917208960 is the second brightest candidate in our catalogue,
marked in bold in Table~\ref{tab:Data} and Fig.~\ref{fig:lcs}.

\begin{figure} 
\centering
{\includegraphics[width=0.5\textwidth]{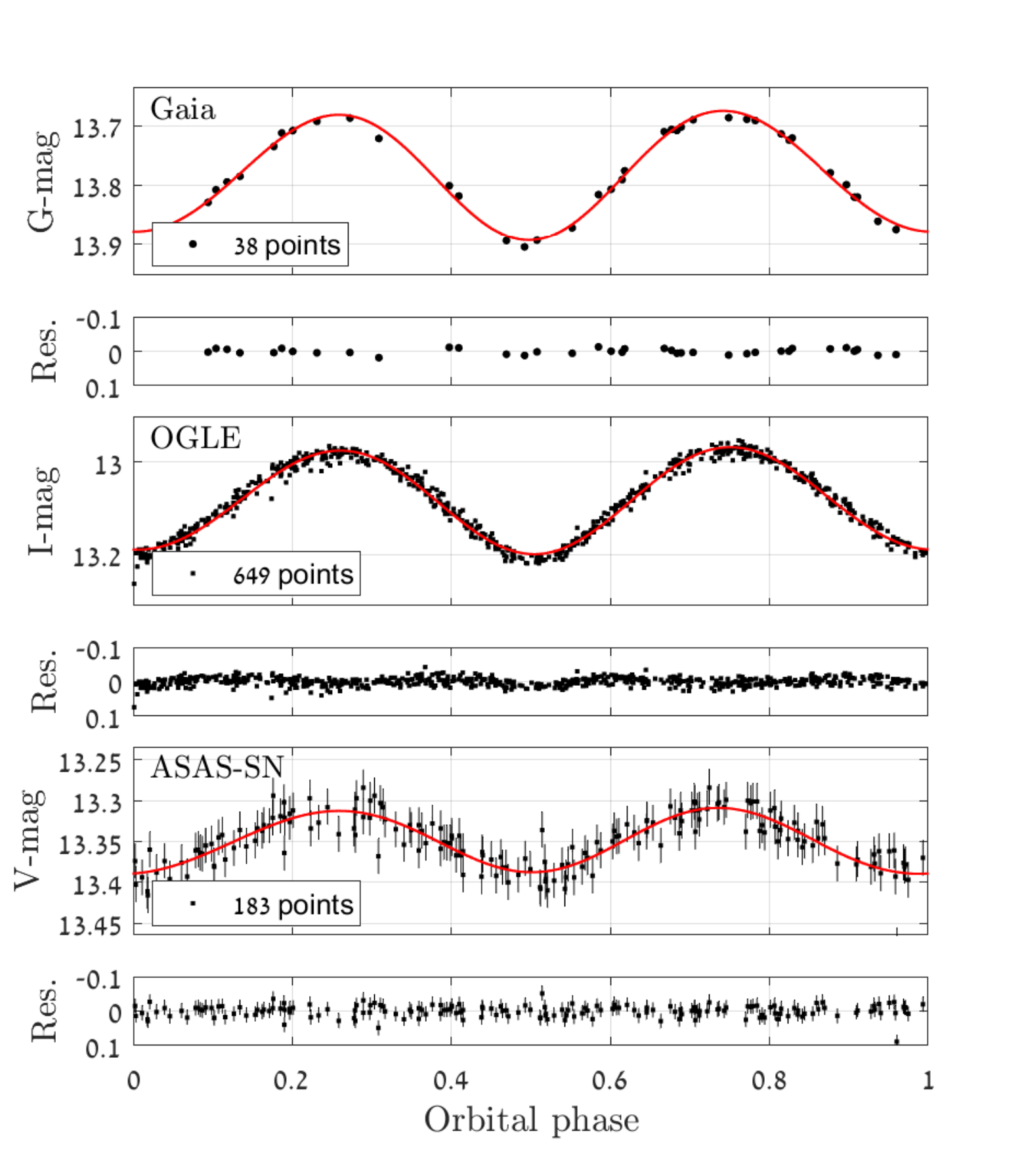}}
\caption{Folded light curves of \gaia DR3 4042390512917208960 in the \gaia G (top panel), OGLE I (middle panel) and ASAS-SN V (bottom panel) bands. The orbital phase is calculated with a period of 0.89522 d and zero phase at BJD 2457581.3996, derived by the \gaia pipeline. A three harmonics model is plotted with a solid line, and the residuals are plotted in the lower panel. The G- and I-band uncertainties, on the order of $1$ ppt, are too small to be noticed.}
	\label{fig:LC_8960}
\end{figure}

\begin{figure} 
\centering
{\includegraphics[width=0.5\textwidth]{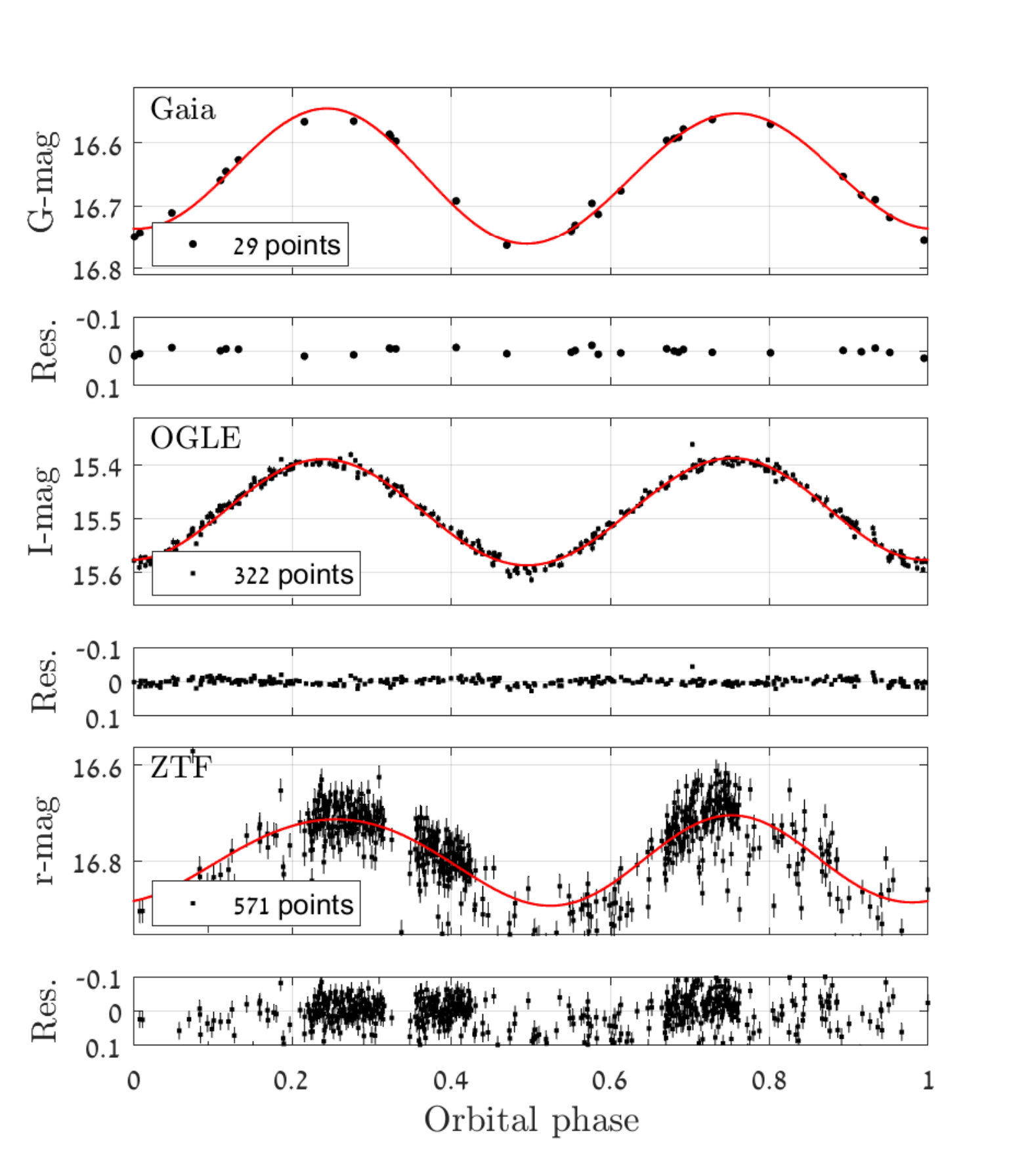}}
\caption{Folded light curves of \gaia DR3 4070409432055253760 in the \gaia G (top panel), OGLE I (middle panel) and ZTF r (bottom panel) bands, as in Figure~\ref{fig:LC_8960}. The orbital phase is calculated with a period of 0.64373 d and zero phase at BJD 2457449.2337, derived by the \gaia pipeline.}
	\label{fig:LC_3760}
\end{figure}

\begin{figure} 
\centering
{\includegraphics[width=0.45\textwidth]{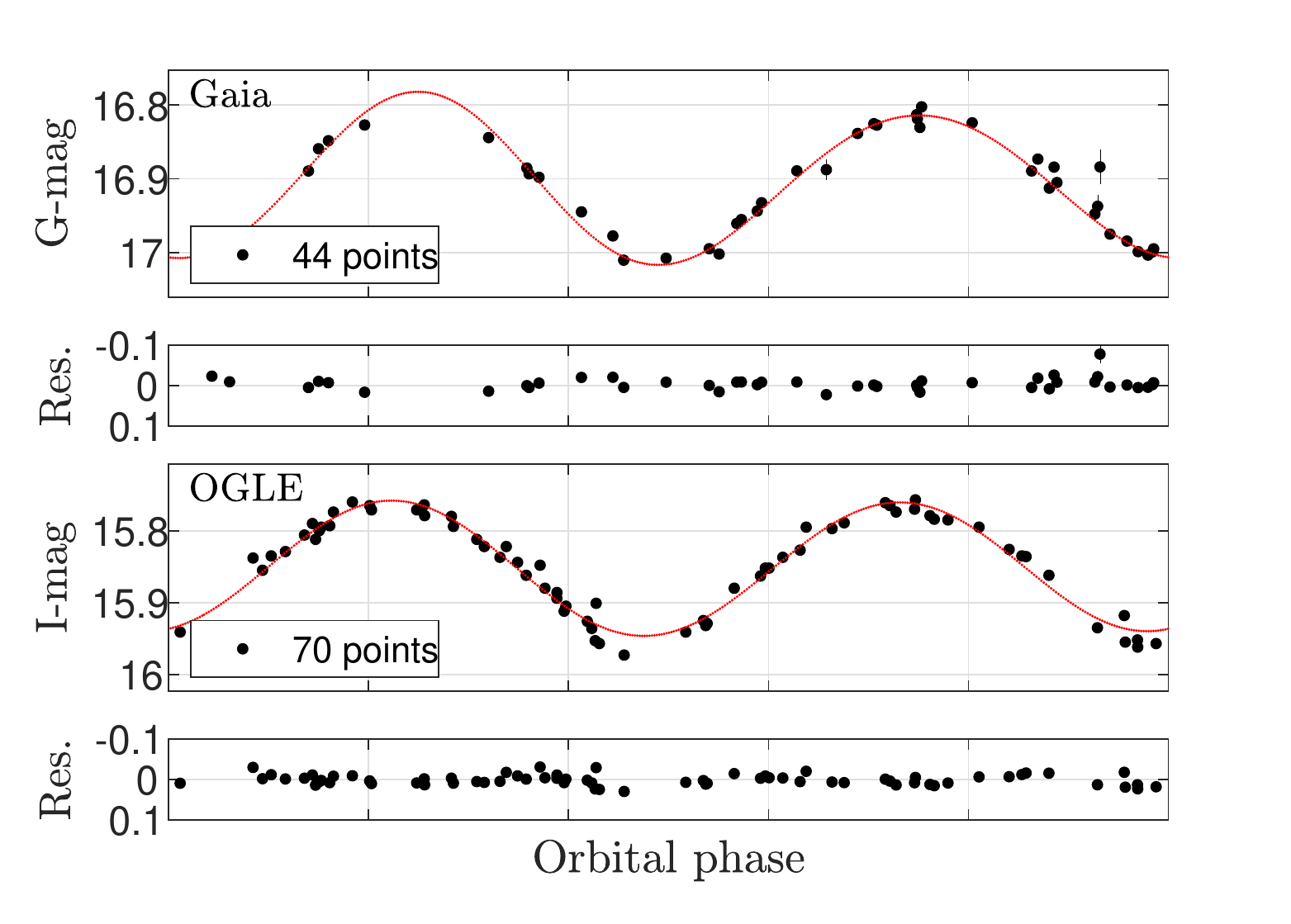}}
\caption{Folded light curves of \gaia DR3 4056017172771375616 in the \gaia G (top panel) and OGLE I (bottom panel) bands, 
as in Figure~\ref{fig:LC_8960}. The orbital phase is calculated with a period of 0.410707 d and zero phase at BJD 2457591.75673, derived by the \gaia pipeline.}
	\label{fig:LC_5616}
\end{figure}

Figures \ref{fig:LC_8960}--\ref{fig:LC_5616} show the available photometry of the three examples, folded with the \gaia period, clearly displaying the \gaia modulation in the different independent data sets.  
Table~\ref{tab:Par} lists some parameters of the photometric analysis of light curves of the three examples --- source identifier, variability classification, second-harmonic Fourier coefficient with its  uncertainty, and the mMMR and its lower-percentile value.


The first star, \gaia DR3 4042390512917208960, was classified as an ellipsoidal variable by the OGLE team, 
but as a contact eclipsing binary (EW in their terminology) by the ASASSN-V group \citep{jayasinghe20}. 
Based on the \gaia and OGLE classification, we do suggest the system is an ellipsoidal variable.
The relatively small amplitude of the ASAS-SN light curve is probably due to light contamination from a neighbouring star, as expected in the dense Galactic Bulge.
The second star,  \gaia DR3 4070409432055253760, was classified as an ellipsoidal variable by the OGLE team, reinforcing our classification.
The two stars have already been analyzed by \citet{gomel21c} in a search for compact companions, deriving similar values for $\hat{q}_{\rm min}^{-1\sigma}$.

The third star,  \gaia DR3 4056017172771375616, with the shortest period of the three systems, was classified as a contact binary (EC in their terminology) by the OGLE team. Although it is quite difficult to distinguish between the two types of modulations, 
the modulation shape is consistent with an ellipsoidal modulation, and follow-up observations are required to determine the correct variability. 


\renewcommand{\arraystretch}{1.0}
\begin{table}
\caption{Parameters of the photometric analysis of light curves of the three candidates
\label{tab:Par}}
\centering 
\begin{adjustbox}{width=0.45\textwidth}
\begin{tabular}{|c|c|c|c|} 
\hline
 & & \thead{$a_{2\mathrm{c}}$} & \thead{$\hat{q}_{\rm min}$} \\ 
& Classification$^1$ & \thead{$a_{\rm 2c,err}$} & \thead{$\hat{q}_{\rm min}^{-1\sigma}$} \\
& & \thead{$\rm [mag]$} & \\
\hline
\hline
$\rm DR3\ 4042390512917208960$  & ELL   & 0.1039    & 2.5   \\
\gmag band                      &       & 0.0022    & 1.8   \\
\hline
OGLE-BLG-ELL-012306             & ELL   & 0.11219   & 2.5   \\ 
I band                          &       & 0.00053   & 1.7   \\
\hline
ASASSN-V J175613.02-335233.3    & EW    & 0.0386          &       \\
V band                          &       & 0.0018          &       \\
\hline 
\hline
$\rm DR3\ 4070409432055253760$  & ELL   & 0.0995    & 2.1   \\
\gmag band                      &       & 0.0027    & 1.5   \\
\hline
OGLE-BLG-ELL-013007             & ELL   & 0.09659   & 1.4   \\ 
I band                          &       & 0.00058   & 1.0   \\
\hline
ZTF 282116300002763             &       &  0.0885          &       \\
r band                          &       &  0.0085         &       \\
\hline 
\hline
$\rm DR3\ 4056017172771375616$  & ELL   & 0.1066    & 2.9   \\
\gmag band                      &       & 0.0031    & 1.9   \\
\hline
OGLE-BLG-ECL-106459             & EC    &  0.0880         &       \\ 
I band                          &       &  0.0024         &       \\
\hline
\end{tabular}
\end{adjustbox}
	\begin{tablenotes}
      \tiny
      \item $^1$ Classification abbreviations represent ellipsoidal variable (ELL), eclipsing contact binary (EC) or contact binary of EW type (EW).
    \end{tablenotes}
\end{table}

Table 6 lists the astrophysical parameters of the three systems, including effective temperature, log gravity, metallicity, radius, mass, and stellar age, as derived by  \citet[\gaia DR3,][]{Fouesneau22}. 
The last column of the table lists the minimum mass of the secondary, derived from mMMR of Table~\ref{tab:Data} and the primary mass. In parentheses 
 the $15.9$ percentile of the mass is given, using the $\hat{q}_{\rm min}^{-1\sigma}$ value. 
The second star does not have an estimate of the astrophysical parameters, and therefore no minimum secondary mass was derived. 

The two stars with radius and mass in Table~\ref{tab:AstroPar} display a radius that indicates a slightly evolved state. In principle, we tried to compose our catalogue of candidates with MS primaries only, to avoid Algol-type binaries, for which the evolved primary can over shine a more massive companion (see below). Nevertheless, the two examples of Table~\ref{tab:AstroPar} are still good candidates, as their assumed secondaries, with minimum mass of $3.2$ and $1.9 M_{\odot}$, are massive enough to show up in the combined luminosity of the systems, if they were MS components.


\renewcommand{\arraystretch}{1.5}
\begin{table*}
\caption{ Astrophysical parameters of the three candidates 
\label{tab:AstroPar}}
\centering 
\begin{adjustbox}{width=0.9\textwidth}
\begin{tabular}{|c|r|r|r|r|r|c|c|} 
\hline
\gaia DR3 & \thead{$\rm T_{eff} \ [K]$} & \thead{$\rm logg$} & \thead{$\rm FeH \ [dex]$} & \thead{$R \ [R_{\odot}]$} & \thead{$M \ [M_{\odot}]$} & \thead{$\rm age \ [Gyr]$} & \thead{$M_{\rm 2, min} \ [M_{\odot}]$} \\ 
\hline
$4042390512917208960$ & $5036^{+442}_{-231}$ & $4.18^{+0.13}_{-0.13}$ & $-0.23^{+0.30}_{-0.77}$ & $3.16^{+0.16}_{-0.13}$ & $1.81^{+0.06}_{-0.05}$ & $1.4^{+0.2}_{-0.2}$ & $4.6 \ (3.2)$ \\ \hline 
$4070409432055253760$ & $4024^{+71}_{-67}$ & $4.80^{+0.05}_{-0.06}$ & $0.10^{+0.05}_{-0.10}$ &  &  & & \\ \hline 
$4056017172771375616$ & $4282^{+52}_{-70}$ & $4.70^{+0.07}_{-0.09}$ & $-0.00^{+0.07}_{-0.10}$ & $1.68^{+0.24}_{-0.16}$ & $0.99^{+0.07}_{-0.05}$ & $11.5^{+1.6}_{-2.2}$ & $2.8 \ (1.9)$ \\ \hline 
 \end{tabular}
\end{adjustbox}
\end{table*}

\section{Discussion}
\label{sec:discussion}

We constructed a catalogue of $6306$ variable stars that might have massive, probably compact, companion in short-period orbit, based on their ellipsoidal modulation. As shown in Fig.~\ref{fig:N-hist}, the period distribution of the candidates peaks at $\sim 0.3$--$0.4$ days, with a moderate decline towards longer periods. 
Obviously, our search is more sensitive at short periods, and therefore the decline is
probably due to selection effects. Out of the whole catalogue we selected $262$ candidates with modified minimum mass ratio larger than unity, with higher probability to have a compact companion. 

As mentioned above, the reality of the candidacy of the binaries in our catalogue, and the sample of stars with large mMMR in particular, depends on two main assumptions:
\begin{itemize}
    \item The observed periodic variability is due to ellipsoidal modulation.
    \item The star is on the MS.
\end{itemize}

As shown, the two assumptions are not always fulfilled. Some systems are  contact binaries (CB), with two components that are not detached.
As pointed out above, it is not straightforward to distinguish between CBs and ellipsoidal variables. 
For example, \gaia DR3~$4042390512917208960$, discussed in Section~\ref{sec:example}, was classified by our pipeline and by  OGLE as an ellipsoidal variable,  but as a contact eclipsing binary by ASAS-SN. 
Nevertheless, 
we suggest that the contamination of the catalogue by contact binaries is not severe, because CBs, by their nature, must have short orbital periods, mostly below $0.25$ d, as shown by \citet{rucinski10}. We have avoided periodic variables with such short periods. 

In other cases, the modulations might be due to single-star variability, like stellar pulsation or rotation. One example is probably 
\gaia DR3~$4068402346632484864$, whose folded light curve resembles an \rrl ab-type modulation, as shown in Figure~\ref{fig:lcs}.
Furthermore, \rrl c-type  variables might  better mimic ellipsoidal variability.
%
However, we suggest that only a few single-star variables are hidden in the catalogue, 
due to the efficiency of the CU7 classifier \citep{DR3-DPACP-165}.

Finally, some of the stars in the catalogue might be real ellipsoidal variables but with a MS companion.
In some cases, the derived mMMR is smaller than unity. We nevertheless included them in the catalogue, because the mMMR can be substantially lower than the actual minimum mass ratio, depending on the actual stellar mass and radius of the primary. In other cases, the primary star is slightly evolved and therefore might outshine a MS secondary. However, as shown in Fig.~\ref{fig:CMD}, a considerable portion of the catalogue stars with extinction estimates are on the main sequence, and even cases with slightly evolved stars, like the two examples discussed in Section~\ref{sec:example}, cannot hide a MS companion.

Nevertheless, despite all cautious measures, many of the candidates of the catalogue may not have a compact-object companion. Therefore, only spectroscopic radial-velocity (RV) follow-up observations can validate the high mass ratio of the candidates. 
However, even before RV resources are devoted to follow-up observations, one can add more photometric measurements to the light curve, to better study the shape of the periodic modulation, at different photometric bands in particular. This is expected in future \gaia data releases,\footnote{https://www.cosmos.esa.int/web/gaia/release} for which the number of FoV transits is going to double, including the \gbp and \grp 
measurements, which were barely used here. One should also try to estimate the mass, radius and temperature of the primary star. These can be obtained either from the stellar position on the CMD, and/or from its spectral energy distribution, when infrared observations and parallax are available. 
Any deviation from the mass-radius-temperature relations of MS stars might indicate that either the primary is not on the MS, or the secondary substantially contributes to the brightness of the system, making the star a less attractive candidate for having a compact companion.

Another indication for a compact companion may come from X-ray observations \citep[e.g., ][]{forman78,voges99,evans10,webb20}. See, for example, the discussion of \gaia DR3 $5966509571940818048$ in Section~\ref{sec:Xray}. Some dormant compact objects might have temporal X-ray emission \citep[e.g.,][]{remillard06,belloni10}, and therefore have not been detected as X-ray sources yet. Therefore, any X-ray survey or pointed observations of the best candidates can be very useful.  

When follow-up RV are performed, only a few measurements with medium precision, on the order of $10$ km/s, should be enough to establish the binarity and the minimum mass ratio of each system, as the presumed orbital period is known, and the expected RV amplitude is on the order of 100 km/s.  
Unfortunately \gaia RVS measurements \citep{katz19} could not have been used for the candidates of our catalogue, because their stellar brightness and/or temperature are not in the RVS effective range. Thus, our candidates necessitate a dedicated program of follow-up observations.

Such RV follow-up project can be done, for example, with multi-object spectrographs like SDSS-V \citep{kollmeier17} and the upcoming
4MOST \citep{deJong19}. The magnitude limit of these spectrographs depends, obviously, on the exposure time, spectral information of the systems and the RV precision required. All in all, we assume that for our purpose the limit is about 19th mag. This limit is marked in Figs.~\ref{fig:Gmag-23K-hist} and \ref{fig:Gmag-hist}, indicating that most of our candidates can be followed by these spectrographs.

The catalogue presented here, although with unknown level of contamination, has the capacity of opening a new window to study short-period binaries with compact-object companion, either a BH or a NS, and sometimes a white dwarf, when some of the candidates are confirmed by additional observations.

\section*{Acknowledgments}
%
We are indebted to the referee, who contributed illuminating comments and suggestions on the previous version of the manuscript, helping us improve substantially the paper.
This research was supported by Grant No. 2016069 of the United States-Israel Binational Science Foundation (BSF) and by Grant No. I-1498-303.7/2019 of the German-Israeli Foundation for Scientific Research and Development (GIF) to TM.
We have made use of data from the ESA space mission Gaia, processed by the Gaia Data Processing and Analysis Consortium (DPAC). Funding for some of the DPAC participants has been provided by Gaia Multilateral Agreement, which include, for Switzerland, the Swiss State Secretariat for Education, Research and Innovation through the ESA Prodex program, the "Mesures d’accompagnement", the "Activit\'es Nationales Compl\'ementaires", the Swiss National Science Foundation, and the Early Postdoc.Mobility fellowship; for Belgium, the BELgian federal Science Policy Office (BELSPO) through PRODEX grants; for Italy, Istituto Nazionale di Astrofisica (INAF) and the Agenzia Spaziale Italiana (ASI) through grants I/037/08/0, I/058/10/0, 2014-025-R.0, and 2014-025-R.1.2015 to INAF (PI M.G. Lattanzi); for France, the Centre National d’Etudes Spatiales (CNES). Part of this research has received funding from the European Research Council (ERC) under the European Union’s Horizon 2020 research and innovation programme (Advanced Grant agreements N670519: MAMSIE "Mixing and Angular Momentum tranSport in MassIvE stars"). This research has made use of NASA’s Astrophysics Data System, the VizieR catalogue access tool, CDS, Strasbourg, France.

\bibliographystyle{aa}
\bibliography{CCE_bib}
\appendix

\end{document}